\documentclass[12pt]{emulateapj} 
\usepackage{natbib}
\usepackage{epsfig} 
\usepackage{times}
\def\be{\begin{eqnarray}}   \def\ee{\end{eqnarray}}

\def\sec#1{Section~\ref{sec:#1}}
\def\fig#1{Figure~\ref{fig:#1}} 
\def\tab#1{Table~\ref{tab:#1}}
\def\equ#1{Equation~(\ref{equ:#1})}

\def\figp#1{Figure~\ref{fig:#1}}

\def\future_paper{S. Sharma et al. 2010, in preparation} 

\newcommand{\kpc}{\>{\rm kpc}} 

\newcommand{\kms}{\>{\rm km}\,{\rm s}^{-1}}

\newcommand{\degree}{\ensuremath{^\circ}}
\begin{document}
\title[Group finding in the stellar halo using M-giants in 2MASS]{Group finding in the stellar halo using M-giants in
2MASS: An extended view of the Pisces Overdensity?}
\author{Sanjib Sharma$^1$, Kathryn V Johnston$^1$, Steven R.  Majewski$^2$,
  Ricardo R. Mu\~noz$^3$, Joleen K. Carlberg$^2$ and James Bullock$^4$}
\affiliation{$^1$ Department of Astronomy, Columbia University, New York, NY-10027
} \affiliation{$^2$ Department of Astronomy, University of
  Virginia, Charlottesville, VA- 22904} \affiliation{$^3$ Department of Astronomy,
  Yale University,New Haven, CT-06520} \affiliation{$^4$ Center for
  Cosmology, Department of Physics \& Cosmology, University of California
  Irvine, CA-92697}
\begin{abstract}
  A density based hierarchical group-finding algorithm is used to identify
  stellar halo structures in a catalog of M-giants from the Two
  Micron All Sky Survey (2MASS). 
The intrinsic brightness of M-giant stars means that this catalog probes 
deep into the halo
  where substructures are expected to be abundant and easy to detect.  
  Our analysis reveals 16 structures at high Galactic
  latitude (greater than $15^{\circ}$), of which 10 have been previously identified. Among the six new
  structures two could plausibly be due to masks applied to the data, one is associated with
  a strong extinction region and one is probably a part of the
  Monoceros ring. Another one  originates at low latitudes, suggesting some contamination from disk
  stars, but also shows protrusions extending to high latitudes, implying
  that it  could be a real feature in the stellar halo.  
  The last remaining structure is free from the defects
  discussed above and hence is very likely a satellite remnant. 
  Although the extinction in the direction of the structure is very
  low, the structure does match a low temperature feature in the
  dust maps. While this casts some doubt on its origin, the low 
  temperature feature could plausibly be due to real dust in the
  structure itself. 
  The angular position and distance of this structure encompass the Pisces
  overdensity traced by RR Lyraes in Stripe 82 of the Sloan Digital Sky Survey (SDSS).
However, the 2MASS M-giants indicate that the structure 
 is much more extended than what is visible with the SDSS, with the point 
  of peak density lying just outside Stripe 82. 
The morphology of the structure is more like a cloud than a stream and reminiscent of that seen in simulations of satellites disrupting along highly eccentric orbits.
 This finding is consistent with expectations of structure formation within the currently favored cosmological model:
assuming the cosmologically-predicted satellite orbit distributions  are correct,  prior work indicates that such clouds should be the dominant debris structures at large Galactocentric radii ($\sim 100$ kpc and beyond). 
\end{abstract}



\section{Introduction} 

Under the currently favored $\Lambda$CDM model of galaxy formation,
the stellar halo is thought to have been built up, at least in part,
hierarchically through mergers of smaller satellite systems.
Signatures of these mergers should be apparent as structures in the
stellar halo
\citep{1998ApJ...495..297J,1999MNRAS.307..495H,2001ApJ...548...33B,2008ApJ...689..936J}.
In recent years observations have lent support to the hierarchical
picture with the discovery of
a number of streams and structures of stars in the stellar halo of the
Milky Way.  The most prominent of these structures are the
tidal tails of the Sagittarius dwarf galaxy
\citep{1994Natur.370..194I,1995MNRAS.277..781I,2003ApJ...599.1082M},
the Virgo overdensity \citep{2008ApJ...673..864J}, the
Triangulum-Andromeda structure
\citep{2004ApJ...615..732R,2004ApJ...615..738M,2007ApJ...668L.123M}
and the low latitude Monoceros ring
\citep{2002ApJ...569..245N,2003ApJ...588..824Y,2005ApJ...626..128P,2005MNRAS.362..906M}.  

The mapping of these low surface brightness structures can be
attributed to the advent of large scale stellar catalogs derived
from surveys such as the Two Micron All Sky Survey (2MASS) and the
Sloan Digital Sky Survey (SDSS).  Typically, a judicious color
selection is applied to objects in a survey in order to maximise the
presence of stars with some well-defined absolute magnitude range.
Structures are then identified by visually inspecting sky-projections of
the stellar density in slices of apparent magnitude.  Future surveys,
such as GAIA \citep{perryman02}, LSST \citep{2009AAS...21346003I}
SkyMapper \citep{2007PASA...24....1K} and PanSTARRS, will explore the
stellar halo to greater depth, with even larger numbers of stars and
in more dimensions and should be sensitive to even more structures.

While discovery by visual inspection has proved successful so far, the
scale and sophistication of the maps generated from these data sets
(both current and future) motivate an exploration of methods that can
instead objectively identify structures.  This task is well suited to
clustering algorithms, which have enjoyed great success in other areas
of astronomy, e.g., identifying galaxy groups in redshift surveys
\citep{2004MNRAS.348..866E} or identifying halos in cosmological
simulations
\citep{2007MNRAS.374....2R,2001MNRAS.321..372J,1993MNRAS.262..627L}.
The stellar halo presents unique challenges for such algorithms.  The
structures in the stellar halo have arbitrary shapes, they span a wide
range of densities that cannot be separated by a single isodensity
contour and they can have nested substructures.
In this paper we present an objective analysis of substructures in the
stellar halo using the code EnLink \citep{2009ApJ...703.1061S}, which
is a density-based hierarchical group finder.  The code is ideally
suited for this application for four reasons.  First, a density-based
group-finder is able to identify irregular groups. Second, EnLink's
clustering scheme can identify groups at all density levels. Third,
EnLink's organizational scheme allows the detection of the full
hierarchy of structures.  Finally, the group finder gives an estimate
of the significance of the groups, so spurious clusters can be
ignored.

Among the existing surveys, the 2MASS catalog of M-giant stars and
the SDSS catalog of F and G type main-sequence-turnoff (MSTO) stars
provide the clearest global views of the stellar halo. While SDSS
contains a larger number of stars than 2MASS M-giants, it covers only
about 10,000 ${\rm deg}^2$ (1/4 of the sky) in area. Moreover, the
magnitude limit of SDSS means that MSTO stars can probe the stellar
halo only out to 35 $\kpc$ \citep{2008ApJ...680..295B} while M-giant
stars in 2MASS probe out to 100 $\kpc$
\citep{2003ApJ...599.1082M}. This implies that the M-giant stars in
2MASS not only cover a factor of about 90 in volume more than the MSTO
stars in SDSS, but also probe the outer halo where the substructures
are expected to be more abundant and have higher density contrast. 
Hence, we choose to apply EnLink to the
2MASS M-giant sample with the aim of objectively identifying
substructures within it. 

Note that using M-giants as tracers also has its share of disadvantages. 
First, M-giants are a rare population so the total size of the 
survey is much smaller than the SDSS MSTO sample.
Second, M-giants are metal rich, intermediate-age stars with metallicity 
[Fe/H] typically greater than $>-1.5$.
Hence, applying a group finder to an M-giant survey will preferentially detect 
high metallicity debris from the few massive recently-accreted objects and
will be insensitive to ancient or low-metallicity debris that originates from the many more low mass progenitors.
The advantage of this bias against ancient or low-metallicity stars is that it will increase the sensitivity to the rare, recent, high-mass events. 
However, building a census of debris from all types of accreting
objects would require combining these results with those from other
surveys---to be discussed in detail in a forthcoming paper (\future_paper).

The paper is organized as follows: \sec{datasets} describes the 2MASS
M-giant data set used in the paper; \sec{methods} discusses the
methods employed for analyzing the data, i.e., group finding; in
\sec{2mass_struct} we describe the structures identified by the
group-finder in the 2MASS M-giant sample; 
and finally, we summarize our
findings in \sec{summary}.

\section{Selecting M-giant halo stars from the 2MASS data} \label{sec:datasets}
\label{sec:preprocess}
The 2MASS all sky point source catalog contains about $471$ million objects
(the majority of which are stars)
with precise astrometric positions on the sky and photometry in three bands
$J, H,$ and $K_s$. The survey catalog is $99\%$ complete for $K_s<14.3$.  
An initial sample of candidate M-giants was
generated by applying the selection criteria: \be
K_s   & < & 14.0 \\
J-K_s & > & 0.85 \\
J-H   & < & 0.561(J-K_s)+0.36 \\
J-H & > & 0.561(J-K_s)+0.19.  \ee 
All magnitudes in the above equations are in
the intrinsic, dereddened 2MASS system (labeled with subscript 0
hereafter), with dereddening applied using the
\cite{1998ApJ...500..525S} extinction maps.  These selection criteria and the
dereddening method are similar to those used by \cite{2003ApJ...599.1082M} to
identify the tidal tails of Sagittarius dwarf galaxy. In general, for
$(J-K_s)_0>0.85$ giants begin to separate from dwarfs in the near-infrared
color-color diagram, with redder colors leading to better discrimination.
However, the number density of giants in the catalog falls off rapidly as a
function of color. As a compromise between quality (i.e.,  the level of
contamination by disk dwarfs) and quantity we restrict our search to stars
with $(J-K_s)_0>0.97$. This generates a list of about $450,000$ stars
spanning a magnitude range of $4.12-14.0$ in the $(K_s)_0$ band.

\begin{figure}
  \centering \includegraphics[width=0.50\textwidth]{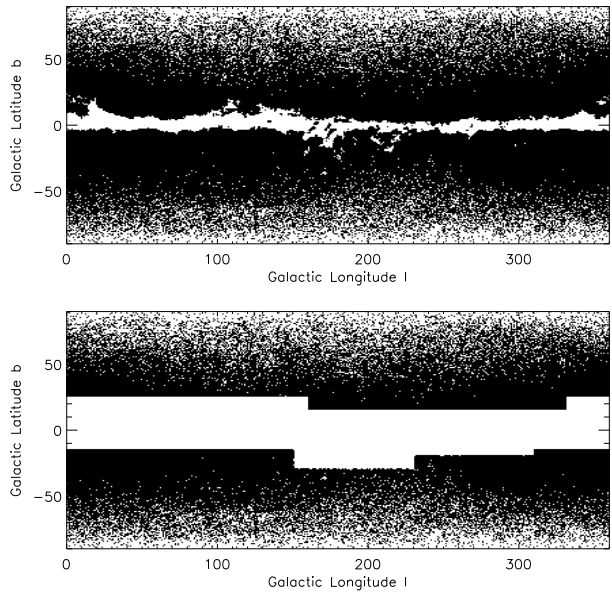}
\caption{ Latitude vs. longitude scatter plot of M-giant stars
  identified in the 2MASS
data. Top: original data containing extinction regions at low
latitudes. 
Bottom: distribution of stars after masking the extinction regions
by means of rectangular patches and retaining stars with latitude 
$b>15^{\circ}$.
\label{fig:f1}}
\end{figure}

\begin{figure}
  \centering \includegraphics[width=0.5\textwidth]{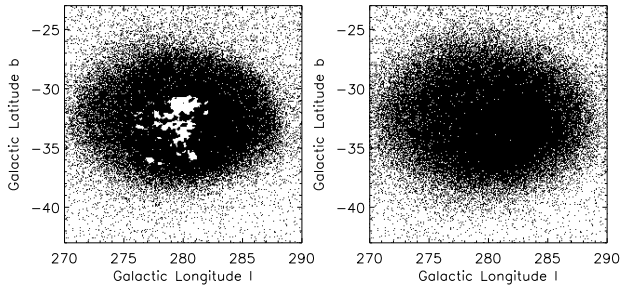}
\caption{Latitude vs. longitude scatter plot of M-giant stars in the LMC
region. Left: original data showing extinction regions. Right: the same region
after adding a dispersion of $1^{\circ}$ to stars satisfying 
$\sqrt{(l-280^{\circ}.0)^2+(b+33^{\circ}.0)^2}<10^{\circ}$.
\label{fig:f2}}
\end{figure}

\begin{figure}
  \centering \includegraphics[width=0.5\textwidth]{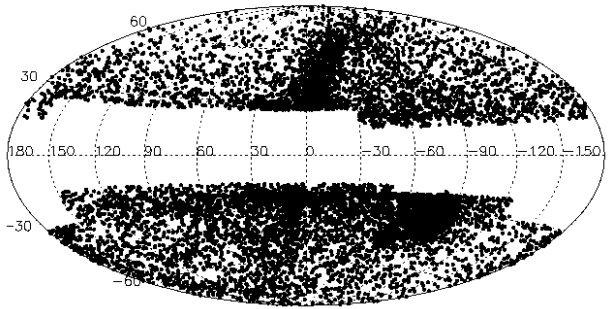}
\caption{ An Aitoff plot in galactic coordinates of the final 2MASS
  M-giant catalog that is used for the group-finding analysis.
\label{fig:f3}}
\end{figure}

Since we are interested in the stellar halo, 
we further refine our selection with geometrical factors aimed at reducing
contamination by foreground disk stars, as well as adopting masks to
cover regions of high dust extinction.
First, we impose the twin requirements that $(K_s)_0>10$ and
$(K_s)_0{\rm sin}(b) > 14.0 {\rm sin}(15^{\circ})$.
The former condition gets rid of stars near the Sun, while the latter
limits the contribution by
stars that are further away, but lie close to the Galactic plane.  
At low latitudes the distribution of stars is
not contiguous owing to the presence of extinction clouds, which in some
regions extend to a latitude of $30^{\circ}$. To avoid identifying spurious
structures and at the same time retain as much low latitude data as possible
we mask the high extinction regions by means of a set of rectangles in $(l,b)$
space, as shown in \fig{f1}.
Finally, there are some extinction holes in the region of the Large Magellanic
Cloud (LMC). We fill these up by identifying the stars lying within a region
defined by $\sqrt{(l-280^{\circ}.0)^2+(b+33^{\circ}.0)^2}<10^{\circ}$ and
adding a dispersion of $1^{\circ}$ to their original latitude and longitude
coordinates, as illustrated in the left and right panels of \fig{f2}.
After applying all of the selection criteria, the final sample contains
$59,392$ stars. An Aitoff plot of these M-giants is shown in \fig{f3}.  

\begin{figure}
  \centering \includegraphics[width=0.5\textwidth]{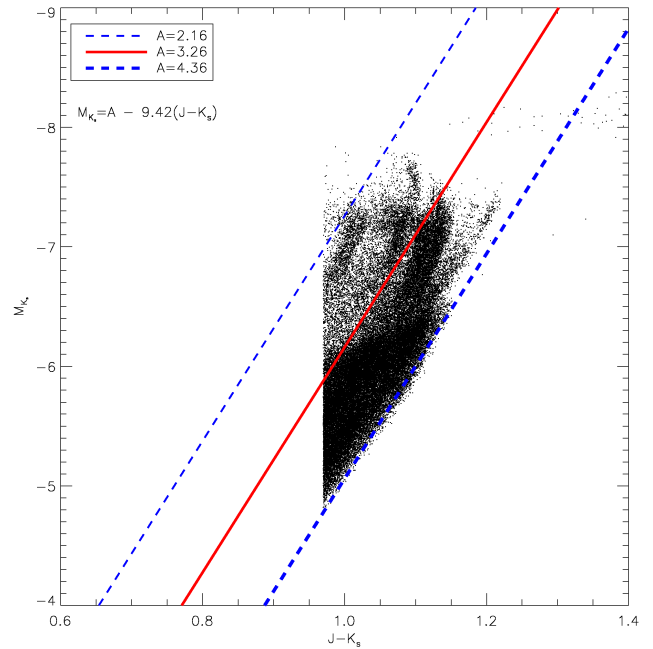}
\caption{ Absolute magnitude of M-giants as a function of its color 
in a $K_s<14$ volume limited sample of  \citet{2005ApJ...635..931B} 
simulated stellar halo (halo-2). The relationship is well represented by a 
function of the form $M_{K_s}=A-9.42(J-K_s)$. 
The solid line with a value of $A=3.26$ is found to roughly bisect the
distribution of points in the plot. The dashed
lines with $A=3.26\pm 1.1$ represent the range of scatter about 
this relationship.  
\label{fig:f4}}
\end{figure}

A particularly useful property of M-giants is that their absolute
magnitude varies approximately linearly with their color and can be
expressed as 
\be
M_{K_s}=A+B(J-K_s)
\label{equ:colmageq}
\ee
A slope of $B=-9.42$ was found to be a good fit, in the regime
$0.97<(J-K_s)_0<1.2$, to a range of theoretical isochrones 
with [Fe/H]$>-1$ and age in range $6-13$ Gyr. The intercept $A$ however 
depends upon the age 
and metallicity. Since we do not know the age and metallicity we
choose to adopt a constant value of $A=3.26$ that  
 roughly bisects the distribution of $M_{K_s}$
versus $(J-K_s)_0$ in the simulated stellar halos of
\citet{2005ApJ...635..931B} \footnote{The simulated halos were
  converted into a synthetic catalog of stars by utilizing isochrones
  from the Padova group \citep{1994AAS..106..275B, 2008AA...482..883M,
  2004AA...415..571B}. A code was developed for this, details of
  which 
will be presented in a forthcoming paper (\future_paper)
}. 
One such halo is shown in
\fig{f4} . The dashed lines with $A=3.26\pm 1.1$ represent the 
range of scatter about this relationship.  
A detailed discussion of 
the impact of our assumption of a constant age and metallicity    
for detecting structures is given \sec{CMReffect}.  

\section{Methods}\label{sec:methods}
\subsection{Group finding}
\label{sec:group_finder}
\begin{figure}
  \centering \includegraphics[width=0.5\textwidth]{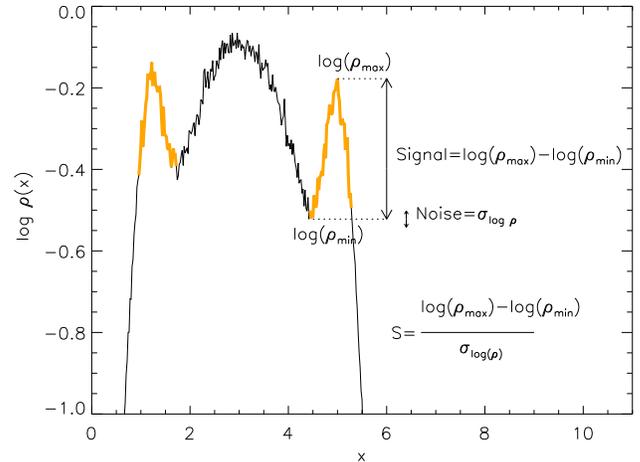}
\caption{Schematic illustration of the group-finding scheme in one dimension.
The plot shows the distribution of density in space for three superposed
Gaussian distributions along with a noise of 0.02 dex. The substructures that 
are bounded by a valley are represented by the thick light gray (orange)
curve.  The maximum and minimum values of the density in a group are used to 
calculate its significance.
\label{fig:f5}}
\end{figure}

In this paper we use the density-based hierarchical group-finder EnLink
\citep[described in detail in][]{2009ApJ...703.1061S} that can cluster a set
of data points defined over an arbitrary space.  For our application the stars
are treated as the data points and the coordinates of the data points are
defined by the position of the stars in three-dimensional space. The group
finding scheme of EnLink is similar to ISODEN \citep{593456} and SUBFIND
\citep{2001MNRAS.328..726S} and is based on the fact that a system having more
than one group will have peaks and valleys in the density distribution, the
peaks being formed at the center of the groups and the valleys or saddle
points where they overlap. The peaks are identified as groups and the region
around each peak, which is bounded by an isodensity contour corresponding to
the density at the valley, is associated with the group. This is shown
schematically in \fig{f5} for a one-dimensional case. The valleys also define
connections between groups and these are used to assign a parent/child
relationship between the groups, resulting in a hierarchy of clusters.

To implement the above scheme EnLink first calculates the density using a
nearest neighbor scheme, where the number of nearest neighbors $k_{\rm
  den}$ is
fixed and is supplied by the user. A list of $k_{\rm link}=10$ nearest neighbors
for each data point is also computed and stored.  Next, the points are sorted
according to their density in descending order and stored in a
list. Starting from the densest, each point from this sorted list is chosen
successively and acted upon according to three options:
\begin{enumerate}
  \renewcommand{\labelenumi}{(\roman{enumi})}
\item If the point does not have a neighbor denser than itself a
  cluster is created and the particle is added to it;
\item If its denser neighbors belong to a unique cluster the particle is also
  added to it;
\item If the denser neighbors belong to different clusters, the two nearest
  clusters are selected and the point is added to the cluster having the
  closest neighbor. Also, the smaller of the two nearest clusters becomes a
  sub-cluster of the larger (now the ``parent'' cluster) and all future
  particles that need to be added to the smaller cluster are added to the
  parent from then on---a process known as sub-cluster attachment.
\end{enumerate}

EnLink employs an additional strategy to screen out spurious groups that
can arise due to Poisson noise in the data.  EnLink defines the significance $S$
for a group as a ratio of signal associated with a group to the noise in the
measurement of this signal (see \figp{f5} for a schematic illustration). 
The contrast $\ln(\rho_{\rm max})-\ln(\rho_{\rm
  min})$ between the peak density of a group ($\rho_{\rm max}$) and valley
($\rho_{\rm min}$) where it overlaps with another group can be thought of as
the signal, and the noise in this signal is given by the variance
$\sigma_{\ln(\rho)}$ associated with the density estimator.  Combining the
definitions of signal and noise then leads to \be S=\frac{\ln(\rho_{\rm
    max})-\ln(\rho_{\rm min})}{\sigma_{\ln \rho}}.
\label{equ:significance}
\ee 
For Poisson-sampled data the distribution of density as estimated by
  the code using the kernel scheme is log-normal and the variance satisfies
  the relation $\sigma_{\ln(\rho)} = \sqrt{V_d ||W||_{2}^2/k_{\rm den}}$,
  where $k_{\rm den}$ is the number of neighbors employed for density
  estimation, $V_d$ the volume of a $d$-dimensional unit hypersphere and
  $||W||_{2}^2$ the $L2$ norm of the kernel function
  \citep{2009ApJ...703.1061S}.  For our case, $d=3$ and $k_{\rm den}=30$ and
  the variance is $\sigma_{\ln(\rho)}=0.22$. 

The distribution of the significance parameter $S$ is close
to a Gaussian function for Poisson-sampled data. This implies that
spurious groups in general have low $S$ and their probability of occurrence
falls off like a Gaussian distribution with increasing $S$. Hence, selecting
groups using a simple threshold in the significance $S_{\rm Th}$ can get rid
of the spurious groups.  EnLink uses this recipe to calculate the significance
of the groups. All groups below $S_{\rm Th}$ are denied the status of a group
and are merged with their respective parent groups.

\subsection{Parameter Choices}
The number and properties of groups recovered by our clustering algorithm
depend in part on the parameters adopted for the group finder itself, as well
as  how the data are transformed from observable to real-space.
In this section, we first define measures to 
evaluate the performance of our clustering scheme (\sec{cluseval})
and subsequently use these measures to guide our choice 
of data transformation (\sec{transformation}) and group-finding parameters (\sec{param}).

\subsubsection{Evaluation of clustering} \label{sec:cluseval}
Let $Q$ be a set of data points with two partitions $I$ and $J$,  
$I$ being the set of intrinsic classes that are known a priori   
and $J$ being the set of groups or clusters found by the group finder. 
In our case the data points are the stars in the halo and the intrinsic 
classes are the individual satellite systems that make up the halo. 
Overlaps between the two partitions are given 
by the contingency matrix $n_{ij}$, which gives the number of data points 
common to both class $i \in I$ and group $j \in J$.  
The class that is most frequent (${\rm arg max}_{i \in I} n_{ij}$) in
a group is the class {\it discovered} by the group, and
$D_i$ is the set of all groups in which class $i$ is discovered.

One measure of success for our group finder is the degree to which recovered
groups represent intrinsic classes, which in our case correspond to real physical associations.
We therefore define {\it purity} as the
fraction of correctly classified points in a group $j$ :
\begin{eqnarray}
{\rm Purity}(j) & = & \frac{{\rm max}_{i \in I}\{n_{ij}\}}{n_{.j}},
\end{eqnarray}
where $n_{.j}=\sum_i n_{ij}$ is the total number of data points
in that group.
The mean value of purity $P=\sum {\rm Purity}(j)/|J|$
is then a good indicator of the overall 
quality of the clustering.

We would also like to know how much of an intrinsic 
class can typically be recovered---in our case this corresponds to 
reconstructing long-dead satellites.
In clustering algorithms, the fraction of correctly classified points  
in a class summed over all groups where the class is discovered is
traditionally known as the {\it recall} of a class. We modify this
definition slightly to also take into account the purity of the
discovered points and define penalized recall as
\begin{eqnarray}
{\rm PRecall}(i) & = &\sum_{j \in D_i}\frac{n_{ij}}{n_{i.}}({\rm Purity}(j)-0.5)2,
\end{eqnarray}
where $n_{i.}=\sum_j n_{ij}$ is the total number of data points
in class $i$. The total value of penalized recall, $\sum {\rm
  PRecall}(i)$, represents 
the mean number of classes discovered by the group finder along 
with a penalty term for classes discovered with purity less than 0.5.
This is a good indicator of the overall amount of clustering.

While mean purity and total penalized recall are sensitive to different
aspects of clustering, in many situations they vary inversely with each other and hence both of them should be taken into account when evaluating clustering success. We do this 
by defining a clustering performance index (CPI), which is given by   
\begin{eqnarray}
{\rm CPI} & = & \sum_i {\rm PRecall}(i)\frac{\sum_j {\rm Purity}(j)}{|J|}. 
\end{eqnarray}
The larger the value of ${\rm CPI}$ the better are the clustering 
results. Typically the value of ${\rm CPI}$ ranges between 0 and $|I|$.    
The maximum value occurs when both the mean purity and total recall have their 
maximum values, which are 1 and $|I|$ respectively. 
In some extreme circumstances, e.g., when the total recall is negative, 
CPI can be negative and the minimum possible value is --$|I|$.

\subsubsection{Choice of coordinate system and metric} \label{sec:transformation}

The efficiency of detecting structures in a data set depends upon the 
choice of the coordinate system in which the data are described and the 
metric (a function of coordinates that defines the distance between
any two points in a space) used to calculate distances.  The
simplest metric is the Euclidean metric---appropriate when all the
dimensions are of the same physical units, such as the Cartesian coordinate system
defined by the $x, y, z$ position of stars in a three-dimensional space.  The
observational data of stars, however, are in a spherical coordinate system given
by the two angular positions on the sky and the radial distance. If the
uncertainty associated with the coordinates is small, the data can be easily
converted to the Cartesian system.  More realistically, the angular
coordinates can be directly measured with very high precision but the radial
distance needs to be estimated indirectly from the properties of the stars and
hence has large uncertainty associated with it. For example, as 
discussed in \citet{2003ApJ...599.1082M} we expect a distance uncertainty 
of about 18\% for the M-giants in our sample, and in this case using the simple Cartesian coordinate system could severely degrade the quality of clustering.

A common solution in cases having large uncertainty in one of the coordinates 
is to perform a dimensionality reduction and analyze the data in a lower
dimensional space---for example, in our case using angular positions alone.
An alternative to ignoring the radial dimension altogether is to redefine the
radial coordinate in a logarithmic scale and then use this modified radial
coordinate to convert the data to a Cartesian system.  The advantage of this
transformation lies in the fact that while the dispersion in radial distance
$r$ increases linearly with $r$, the dispersion in modified radial coordinate
$\log(r)$ is constant.  This motivates a transformation of our radial
coordinate to $r'=5(\log(r/(10{\rm pc})))-\mu_0$ where $\mu_{0}$ is
a constant that determines the degree to
which the radial dimension is ignored or used. If $\mu_0$ is small the data lie
in a thin shell, which is equivalent to ignoring the radial dimension
altogether. On the other hand, if $\mu_0$ is large the radial dimension is given
more prominence.

In order to demonstrate the effectiveness of our coordinate
transformation we applied the group-finder EnLink (with parameters $k_{\rm den}=30$ and
$S_{\rm Th}=4.25$) to a synthetic stellar halo survey generated from the 
simulations of \citet{2005ApJ...635..931B}.
As a particularly stringent test we chose to look at a stellar halo
that had been constructed entirely from low-luminosity satellites and
hence contained numerous small, low-contrast structures rather than a few large ones
\citep[corresponding to the ``low-luminosity halo'', 
amongst the six non $\Lambda$CDM halo models described in][]
{2008ApJ...689..936J}. A color limit of $0.1
<g-r<0.3$ and a magnitude limit of $M_r<24.5$ (in the SDSS $ugriz$ band)
were used to generate the model halo.
 Two samples were generated from the model,
both with and without 
distance errors---referred to as data $T$ and data $T_{\rm error}$ respectively. The group finder was run in
both the normal coordinate system and the modified coordinate system
(with $\mu_0=8$). 
For data  $T_{\rm error}$ we assumed a distance uncertainty of $\sigma_r/r=0.25$. 
To compare clustering we use two measures: the 
number of detected groups $G$ and the clustering
performance index CPI (defined in \sec{cluseval}). 
The results are tabulated in \tab{tb3}.  It can be seen that 
for data without errors the clustering results are similar in both 
the coordinate systems, but for data 
with errors, clustering is better in the modified coordinate
system as evidenced by the increase in both $G$ and CPI.

Next to choose an appropriate value of $\mu_0$ we compared the
clustering results, for the data $T_{\rm error}$, in the modified
coordinate system with different values of $\mu_0$.
The CPI was found  to be maximum at $\mu_0 \sim
8$ and hence we adopt this value for rest of our analysis. 
It should be noted that the clustering results were not strongly sensitive 
to the exact choice of $\mu_0$, in fact CPI was found to vary very 
little in the range $-10<\mu_0<10$.   
In general, decreasing $\mu_0$ was found to increase the
number of detected groups.
However, the mean purity of groups was found to decrease with $\mu_0$,
so choosing a value of $\mu_0$ too small would mean greater
contamination by spurious groups.

\begin{table}
  \centering
  \caption{\label{tab:tb3} Comparison of Clustering Performance}
  \scriptsize
  \begin{tabular}{lccccc}
    \hline
    Data & Radial Coordinate& $\sigma_r/r$  & Sample Size  & Groups & $CPI$ \\ 
    \hline
    $T$ & $r$ & 0.0 & $2 \times 10^7$ &114 & 11.1 \\
    $T_{\rm error}$ & $r$ & 0.25 & $2 \times 10^7$ & 15 & 0.5 \\
    $T$ & $5\log(r/10 {\rm pc})-\mu_0$ & 0.0& $2 \times 10^7$ & 118 & 8.9 \\
    $T_{\rm error}$&$5\log(r/10 {\rm pc})-\mu_0$ & 0.25& $2 \times 10^7$ & 35 & 2.3 \\
    $T'_{\rm M-giants}$&$5\log(r/10 {\rm pc})-\mu_0$ & 0.15 & $2.5 \times 10^4$ & 15 &  \\
    $T_{\rm M-giants}$&$5\log(r/10 {\rm pc})-\mu_0$ & Eq-5& $2.5 \times 10^4$ & 17 &  \\
    \hline
  \end{tabular}
\end{table}

\subsubsection{Optimum choice of group-finding parameters} \label{sec:param}
\label{sec:optimum}
The two free parameters in the group-finder are the number of neighbors
employed for density estimation, $k_{\rm den}$, and the significance
threshold, $S_{\rm Th}$, of the groups. We select $k_{\rm den}=30$: a smaller
value makes the results of the clustering algorithm sensitive to noise in the
data, while a larger value means that small structures go undetected.  

The choice of the second free parameter, $S_{\rm Th}$, is governed by the desire 
to make the expected number of spurious groups, which can arise due to
Poisson noise in the data, either constant or zero. This is important if 
one wants to reliably use the number of detected groups as a measure of clustering strength.
For a
$d$-dimensional data consisting of $N$ points an optimum value of $S_{\rm Th}$
can be chosen by considering the number of spurious groups with significance
greater than $S_{\rm Th}$ expected for a Poisson-distributed data (i.e., data points being distributed in a finite region of space uniformly 
but randomly). The required expression is given by 
\be
G(>S_{\rm Th})=(1-{\rm erf}(S_{\rm
  Th}/\sqrt{2}))\frac{15.5N}{d^{2.1}k_{\rm den}^{1.2}}
\label{equ:GS}
\ee 
\citep{2009ApJ...703.1061S}.  Since the presence of even one or two
spurious groups can severely contaminate the analysis of structures
we calculate the optimum value of $S_{\rm Th}$ for a given $N$ 
by setting  the expected number of spurious groups 
$G(>S_{\rm Th})=0.5$ in equation (\ref{equ:GS})
and solve for $S_{\rm Th}$.
Using this method we find $S_{\rm Th}=3.75$ for $N=10^5$ (typical
size of the data analyzed in this paper).
\footnote{Note that the significance of a real group in a given data set also increases
with an increase in $N$, primarily due to the improved spatial resolution and
secondarily due to the nature of Poisson noise.}
In general, decreasing $S_{\rm Th}$ decreases the number of recovered groups 
and the value of total recall, but increases the mean purity. 
On the other hand increasing $S_{\rm Th}$ has exactly the opposite behavior. 
This suggests that CPI should be maximum at some optimum value of $S_{\rm Th}$.
In our tests on synthetic halos we do see this
behavior, i.e., for values of $S_{\rm Th}$ for which $G(>S_{\rm Th})=0.5$, CPI also tends 
to be maximum.

\begin{figure}
  \centering \includegraphics[width=0.5\textwidth]{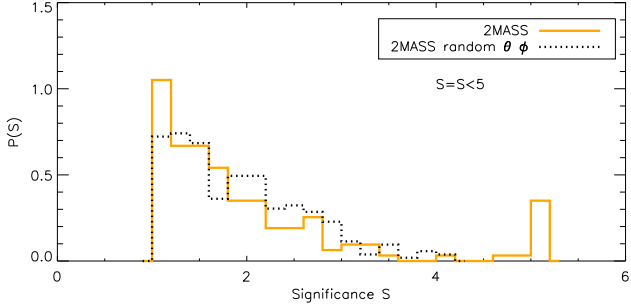}
\caption{ Distribution of significance $S$ for the groups identified by the
group-finder. Groups with $S>5$ were assigned a value of $S=5$.
The plot shows the results for the 2MASS M-giant data and a randomized 2MASS
M-giant data created by choosing the latitude and longitude at random so as to
have a uniform distribution over a sphere. 
\label{fig:f6}}
\end{figure}

As a final confirmation of our choice of threshold 
$S_{\rm Th}$, we generated a data set that
contained only noise by replacing the latitude and longitude measured for
2MASS M-giant stars with values selected at random from a uniform distribution over a
sphere but excluding the low latitude regions (as in the case of the real
2MASS M-giant sample).  We then applied the group-finder to this randomized data-set
with $S_{\rm Th}=1$. The distribution of significance $S$ for the recovered
groups is shown as the dotted histogram in the top panel of \fig{f6}. The
groups have a distribution that is like the tail of a Gaussian, with very few
groups having $S>3.75$. The distribution of groups recovered from the real
2MASS M-giant sample (solid line) is similar to that for randomized data for $S<3.75$.
However, for $S>3.75$ several extra groups can be seen. 
This suggests that choosing a
significance threshold of $S_{\rm Th}=3.75$ to identify groups in a survey
containing $10^5$ points will minimize contamination by spurious groups.

\subsection{Impact of the assumption of a single age and metallicity}\label{sec:CMReffect}
In \sec{datasets} we had tentatively assumed a value of $A=3.26$ in
the color magnitude relation (represented by
\equ{colmageq}, referred to as CMR hereafter), which corresponds
to assuming a single age and metallicity for all the stars.
We now revisit this issue and study the impact of this assumption for 
group-finding studies.

First, we note that as a consequence of working in a space of 
modified radial coordinate (see \sec{transformation}), there is a complete 
degeneracy between the choice of parameters $\mu_0$ and $A$. 
Since our analysis in \sec{transformation} has 
already shown that group finding is insensitive to the exact
choice of $\mu_0$, the same applies for $A$. The relative 
insensitivity of clustering to $\mu_{0}$ or $A$ is because 
a change in value of either of them leads to a mere translation of the data 
in the radial direction while the geometry of structures within the
data remains almost intact.

Although the  ability to identify structures is not sensitive to the
exact choice of $A$, it is sensitive to the scatter of the 
stars about the adopted CMR (as shown in \fig{f4}).  
The standard deviation of distance modulus $\sigma_{\mu}$ computed 
using the adopted CMR  for the full halo was found to be 0.51. 
This high value of $\sigma_\mu$ is
mostly due to systematic differences in metallicities and ages between satellite 
system rather than large ranges internal to each system.
These systematic differences
simply translate
the structures relative to each other in space, 
an effect which does not significantly hamper how well they can be detected. 
For the purpose of detecting structures what matters most is the 
$\sigma_{\mu}$ for individual satellite systems. 
Using the simulated stellar halos of
\citet{2005ApJ...635..931B} we found the mean value of  $\sigma_{\mu}$
for individual satellite systems to be 0.34, i.e., 
distance uncertainty $\sigma_r/r=0.15$, in accordance with our
expectation.  
These dispersion estimates are also in agreement with the results  
of \citet{2003ApJ...599.1082M}, where they report 
$\sigma_{\mu}=0.36$ for the 2MASS M-giants in the core of
Sagittarius. 

Our previous discussion suggests that using the 2MASS M-giants 
along with our adopted CMR for distance determination should 
be roughly equivalent to using a data set with about 15\% dispersion in
distance estimates.
To test this we employ the same low
luminosity halo that was used in \sec{transformation} but now generate
a sample of M-giants using the color magnitude limits  as
described in \sec{datasets} for the real 2MASS M-giant data. 
Equation (\ref{equ:GS}) was used to select
the optimum $S_{\rm Th}$ relevant for the present data size 
and the group finder was run once with $15\%$
errors in distance (data $T'_{\rm M-giants}$) and once with distance 
computed using \equ{colmageq} (data $T_{\rm M-giants}$). The
results are tabulated in \tab{tb3}.
It can be seen that both data sets give 
nearly the same number of groups which demonstrates that for the purpose
of detecting groups, the effect of using a constant age and metallicity 
is similar to that of  data with 15\% error in radial distances.

Comparing the number of detected groups in \tab{tb3} 
for different data sets also allows us to compare the overall 
group-finding efficiency of different schemes. 
We find that of the groups that 
could have been detected without any distance 
errors (data set $T$), only 30\% are detected by data 
$T_{\rm error}$ and 15\% by data $T_{\rm M-giants}$. 
Although the drop in ability to detect groups is quite dramatic,  
it is mainly a reflection of the fact  that 
fainter structures that are harder to detect are 
much more numerous than the brighter and easily detectable structures.
Additionally, our results are biased by the fact that we use 
a hypothetical halo dominated by low mass accretion events
which are also the ones that are preferentially missed 
in a data with measurement errors. Hence, for a realistic 
$\Lambda$CDM halo we expect the percentage of detected groups 
to be slightly higher.

Next, we compare the number of detected groups 
for data $T_{\rm M-giants}$ with data $T_{\rm error}$.
Although the distance error for data $T_{\rm M-giants}$  
is less than that for data $T_{\rm error}$, the number of detected groups
is still about a factor of 2 lower. Three factors could be responsible for this.
First, the sample size
for $T_{\rm M-giants}$ is three order of magnitude lower than
that for data $T_{\rm error}$, which means that the data $T_{\rm M-giants}$
has lower spatial resolution and this makes identification of groups
difficult. Second, M-giant data is biased toward detecting high
metallicity, intermediate-age  stars and would miss low metallicity systems or those accreted long ago, which
dominate by number. Finally, the high mass systems, which are
preferentially sampled by M-giants due to their high metal content, 
are also the most phase 
mixed ones and contribute more to the smooth background, 
making structure detection even more difficult. 
In fact, in a forthcoming paper (\future_paper) 
we demonstrate that, 
despite the low number of stars, a 2MASS type survey can recover most
of the structures that originate from high mass progenitors and are on
orbits of low eccentricity.

\begin{figure*}
  \centering \includegraphics[width=0.72\textwidth]{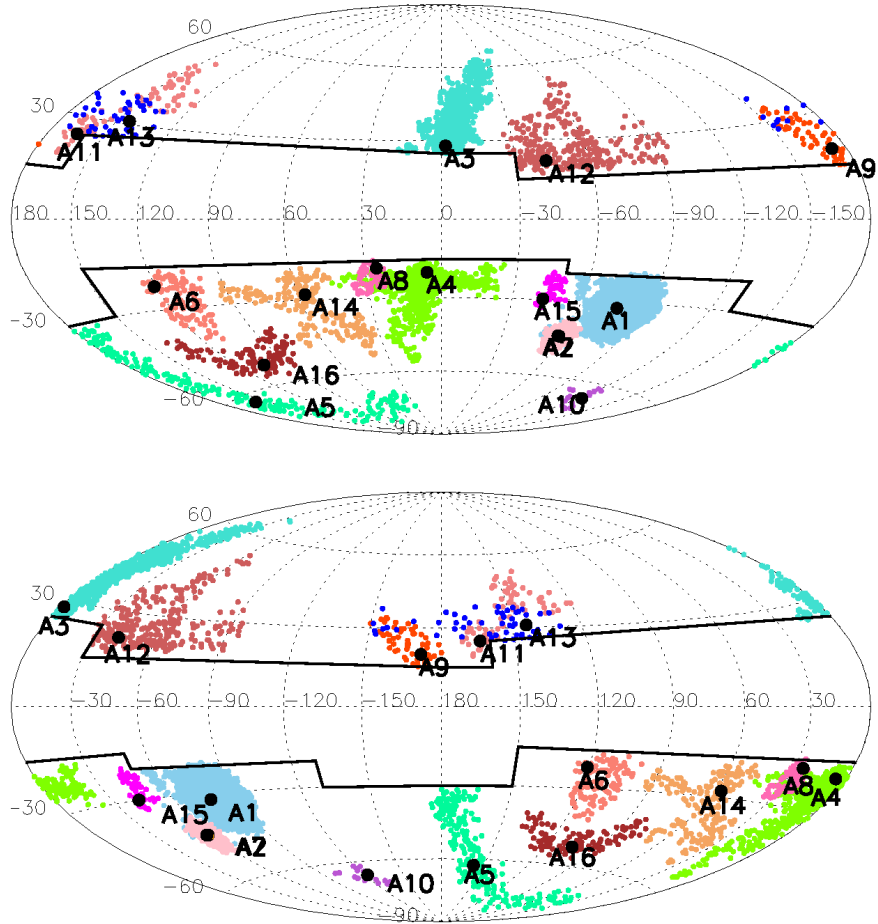}
\caption{Groups found in the 2MASS M-giant sample shown in Aitoff projection
maps centered at $0\degree$ longitude (upper panel) and $180\degree$
longitude (lower panel). Large black filled circles mark the position of the 
density peak in a group. Stars in each group are color coded with a 
unique color and are shown as small filled circles. The solid black
lines mark the low latitude area that is excluded from the analysis.
Note that group A7 lies on the top of group A2.
\label{fig:f7}}
\end{figure*}

\begin{figure}
  \centering \includegraphics[width=0.5\textwidth]{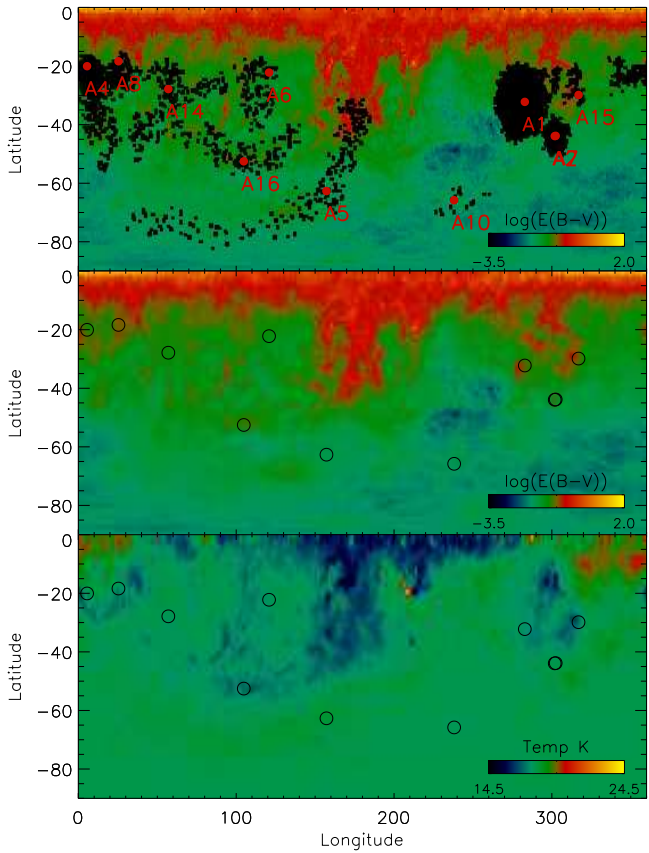}
\caption{Comparison of detected structures with features in dust
  infrared emission maps of \citet{1998ApJ...500..525S}. The 
  panels show the distribution of dust extinction (top two panels) and
  dust color temperature (lower panel), as a function of galactic
  longitude and latitude in the southern hemisphere. The location of
  the structures are marked as circles on the plots. The top two
  panels are the same except for the fact that in the top panel, 
  stars associated with the structures in the 2MASS M-giant sample
  are overplotted. It can be seen that 
  structures A15 and A16  are associated with features both in the 
  extinction and temperature maps.
\label{fig:f8}}
\end{figure}

\begin{figure}
  \centering \includegraphics[width=0.5\textwidth]{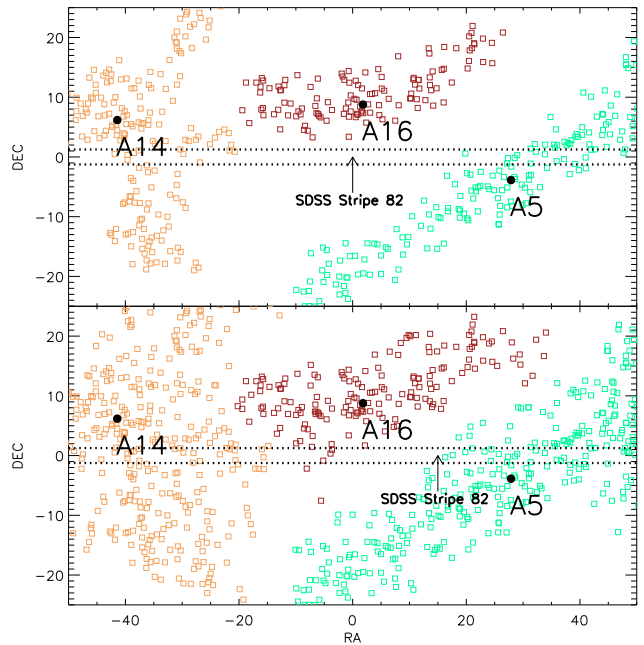}
\caption{Plot in R.A. and decl. of groups found in the 2MASS M-giant
  sample that lie near the SDSS Stripe 82. The filled circles denote
  the point of peak density while the squares denote the data points
  belonging to the groups. The bottom panel shows all possible points
  that can be associated with the density peak of a
  structure by tracing the particles along the direction of density 
  gradients and have density greater than 0.175 times the peak density of the
  structure.   
\label{fig:f9}}
\end{figure}

\section{Results: Structures traced by M-giants in the 2MASS}
\label{sec:2mass_struct}
\begin{table*}
  \centering
  \caption{\label{tab:tb2}Summary of Groups Found in the 2MASS M-giant
    Sample. }
  \small{
    \begin{tabular}{lccccccc}
      \hline 
      Name & Description &$l$ & $b$ & $n_{\rm stars}$ & Sig $S$ & $\rho_{\rm
        peak}$ & Distance\footnote{distance limits computed assuming a
      scatter of 1.1 mag in distance modulus}
       \\
        &  &  & &  &   &       ($\kpc$) \\
      \hline 
      A1 & LMC &       282.865 &      -32.231 &        49234 &       52.9 &      
      2.7 $\times 10^4$ &       $60.1 \pm 30$\\ 
      A2 &  SMC &       301.823 &      -43.925 &         4001 &       33.4 &      
      3.0 $\times 10^4$ &       $64.0 \pm 32$\\ 
      A3 & Sag leading arm, north &       358.130 & 27.985 &      3245 &       27.5 &
      6.5 $\times 10^1$ &       $63.1\pm 32$ \\ 
      A4 & Sag core &       5.51100 &      -20.053 &     1460 &       24.4 &      
      1.8 $\times 10^3$ &       $37.2\pm 19$ \\ 
      A5 &  Sag trailing arm, south &       157.190 & -62.682 &     226 &       4.82 &
      1.0 $\times 10^1$ &       $37.2\pm 19$ \\ 
      A6 & Andromeda &       120.819 &      -22.212 &     117 &       4.49 &     
      6.5 &       $122.0\pm 61$\footnote{The actual distance of
        Andromeda is around 778 kpc and this is much higher than what we
        have derived using M-giants. This discrepancy is because the detected M-giants from
      Andromeda are very rare and bright giants which do not fall on
      the color magnitude relationship that we assume for calculating distances.} \\ 
      A7 & Group in SMC &       302.436 &      -43.837 &   83 &       5.13 &  
      1.7 $\times 10^4$ &       $48.6\pm 24$ \\ 
      A8 & NGC 6822 &       25.393 &      -18.378 &        78 &       4.74 &      
      1.5 $\times 10^1$ &       $92.6\pm 46$ \\ 
      A9 &  Sag trailing arm, south &       187.953 &  19.882 &  64 &    4.54 &      
      2.9 &       $96.6\pm 48$ \\ 
      A10 &  Fornax dwarf Sph &       238.091 &      -65.798 &  39 &    7.58 &      
      7.3 &       $121.3\pm 60$ \\
      \hline
      A11 & Near mask &       164.086 &       24.992 &           79 &     5.18 &      
      3.4 &       $88.2 \pm 44$ \\ 
      A12 & Probably Monoceros ring &       317.865 &       21.908 &          307 &     5.40 &      
      7.5 &       $21.8 \pm 11$ \\ 
      A13 & Near mask &       143.738 &       30.936 &           54 &      3.93 &       
      2.1 &       $22.6 \pm 11$ \\
      A14 & Has protrusions to high $b$ &       56.9910 &      -27.865 &          203 &       5.23 &      
      8.9 &       $97.7 \pm 48$ \\ 
      A15 & Near a strong extinction region & 316.906 &   -29.868 &           76 &       4.99 &      
      2.7 $\times 10^1$ &       $98.6 \pm 49$ \\ 
      A16 & In Pisces constellation & 104.793 & -52.535 & 126 & 6.25 &      
      9.9 &       $102.9 \pm 51$ \\
      \hline 
    \end{tabular}
  }
\end{table*}
Applying our group-finder with $k_{\rm den}=30$ and $S_{\rm Th}=3.75$ to the
2MASS M-giant sample set reveals 16 groups. An Aitoff plot of the groups is
shown in \fig{f7} where each identified group is coded with a unique color and
the filled circles mark the position of the densest particle in the group. A
summary of the group properties is shown in \tab{tb2}. 
Listed in the table are the name of the groups, the galactic latitude
and longitude of the density peak in the groups, the number of stars
in the groups, the significance parameter of the groups, the value of 
peak density and the radial distance of the groups. The first 10 groups
listed in the table can be associated with known structures in the Local
Group, while the other six are new candidate structures.

Among the known structures that have been identified by the group-finder the
densest and most prominent are bound satellite systems such as the
Magellanic Clouds (LMC and SMC\footnote{Group A7 is a subgroup
  embedded within group A2, which is the SMC, hence we consider A7 as a part
  of SMC}), and the core of the Sagittarius dwarf galaxy. Unbound debris in
the form of the streams from Sagittarius are traced beautifully by means of
the structures A3, A5 and A9. Galaxies in the Local Group, like the Andromeda
galaxy, NGC 6822 and the Fornax dwarf spheroidal galaxy, which contribute as
little as 30-100 stars in the sample, are also re-discovered. These findings
both demonstrate the success of the group-finding scheme and lend credibility
to the newly discovered structures.

While our group-finding technique has been successful at revealing 
some of the known structures, others are missing: e.g,. the Virgo overdensity 
\citep{2008ApJ...673..864J}, the Virgo stellar stream 
\citep{2001ApJ...554L..33V,2006ApJ...636L..97D}, the Canis Major dwarf
galaxy \citep{2004MNRAS.348...12M,2005ApJ...633..205M} and the
Hercules Aquila cloud \citep{2007ApJ...657L..89B}. 
The absence of these structures can be understood in terms 
of our sample selection criteria:
the Canis Major overdensity is 
at low latitude ($b=-7\degree.99$) and hence outside the region
explored in this paper; the Virgo stellar stream is metal poor 
$[{\rm Fe/H}]=-1.86$ as suggested by \citet{2006ApJ...636L..97D} and hence 
is most likely not sampled by the metal rich M-giants; 
the Virgo overdensity  is close
to the Sun (6--20 $\kpc$) and largely excluded by 
our selection criteria of $K_s>10.0$  (i.e., distance greater 
than about 15.0 kpc);
the Hercules Aquila cloud is also  nearby (10-20 kpc) and , moreover, the part of it in the northern
hemisphere is centered at $(l,b)=(30\degree,20\degree)$,  
which is outside the region explored by us. 

Next we investigate 
the six newly discovered structures in \tab{tb2}. These could have a
real physical association with 
satellite remnants or they could be artificial overdensities created by dust
extinction regions, masks, contamination from disk stars or Poisson noise. 
For example, structures A11, A12,
A13 and A14 are all at low latitudes and hence possibly 
associated with disk. 
In the cases of structures A11 and A13, both  lie right at the 
edge of one of the rectangular masks (see \fig{f1} and \fig{f3}) 
and this further undermines their authenticity. 
Structure A12 is elongated along the disk, is nearby (distance
of about $23 \kpc$ ) and 
its location matches that reported by \citet{2003ApJ...594L.115R}
for the previously-identified Monoceros ring 
 \citep[see also][]{2003ApJ...588..824Y,2002ApJ...569..245N}.  
Structure A14 also originates at low latitude, but shows a 
protrusion extending to high latitudes that suggests it could be a
real halo structure.  

A comparison of the location of the remaining structures (A15 and A16)  with 
the \citet{1998ApJ...500..525S} infrared dust emission maps
(\fig{f8}) shows that both are associated with high 
extinction and low temperature features in the maps. 
The extinction around structure A15 was found to be particularly high 
while that around A16 is only mildly elevated ($E(B-V)\sim 0.114$ mag). 
Additionally, A16 is at high latitude and also not close to any 
masking region and this makes it a promising new structure
that could correspond to a satellite accretion event. 
The close association with a low temperature feature in the dust 
could either mean that the structure is an artifact of the extinction 
corrections (improperly 
deredenned stars getting spuriously included or excluded in our sample) 
or that the dust feature is associated with 
the real gas and dust in the structure itself. 

Moreover, a structure at a similar location, named the
Pisces overdensity, has recently been discovered in a sample of RR Lyrae
stars in the SDSS Stripe 82 \citep{2009MNRAS.398.1757W,2007AJ....134.2236S}.  
The overdensity has also been spectroscopically confirmed by 
\cite{2009ApJ...705L.158K} using a sample of eight RR Lyraes from SDSS.  
They speculate it to be a bound satellite system
based on the observed velocity dispersion of five of their stars 
being small ($6 \kms$), but at the
same time do not rule out the possibility that it is an unbound satellite 
system due to the large angular width of the overall structure.

To investigate the correspondence of A16 to the Pisces overdensity we 
plot the groups identified in our 2MASS M-giant sample 
alongside  the SDSS Stripe 82, in the top panel of \fig{f9}.
Specifically, the Pisces overdensity has been reported to lie in the
interval  $-25^{\degree}<$R.A.$<0^{\degree}$, with the peak
concentration being at R.A.$\sim -5^{\degree}$ and
at a distance of $r=79.9\pm13.9 \kpc$ \citep[as estimated in][]{2009MNRAS.398.1757W}.
The M-giants in structure A16 are very close to this peak along the
boundary of the strip and at a similar distance given the 
high range of uncertainty ($r=103 \kpc$ with a
range of $\pm 51$)---the offset in distance could either be due to
our arbitrarily adopted value of metallicity in calculating the 
distances to our stars, or to a dramatic distance gradient across the field.

Given these similarities, it is striking that the upper panel of \fig{f9} does
not show any  M-giants from A16 actually in Stripe 82.
The most likely explanation for this lack of stars is 
that the number density of the M-giants associated with the
A16 is 
relatively low within the stripe.
Indeed, a comparison of the number of Sagittarius 
M-giants (group A5)  with that of Sagittarius RR Lyraes in Stripe 82
\citep{2009MNRAS.398.1757W} 
suggests the number of M-giants is probably a factor of 3 lower,
implying that the density peak found in RR Lyraes with SDSS should
contain only a few M-giants. 
Hence, the number density of A16 could decrease 
sufficiently toward the
stripe that it is cut off by the default criteria in the group finder itself,
which truncates a group whenever it intersects a neighboring group.
If we instead relax the default truncation criteria to  also include
points that 
converge to the point of peak density in A16 by following the path along 
local density gradients (i.e., densest nearest neighbor links), we 
find plausible extensions to the group. 
The  bottom panel of \fig{f9} plots the positions for this extended group
with the minimum density threshold of points being set to 0.175 times the  
maximum density within a group. Nearby groups extended with the same
criteria are also shown alongside. It can  be seen that the extended
portion of A16 now matches the distribution of Pisces overdensity RR
Lyraes in Stripe 82.

We also note that the  metallicity of Pisces overdensity has been
reported to be $[{\rm Fe/H}]=-1.5$ by \citet{2009MNRAS.398.1757W}, which 
means that it would be almost undetectable by M-giants. 
But at the same time stars in a satellite system 
do span a range of metallicities, and M-giants could very well 
be sampling the high metallicity stars in the system.

If A16 and Pisces overdensity are related then A16 not 
only offers an independent confirmation of the Pisces overdensity, but also
provides an extended view of it. 
In fact our results show that the point of peak density is located at 
(R.A., decl.)$=(1^{\circ}.81,8^{\circ}.77^)$, which is just outside the range of 
Stripe 82 in SDSS (by about $7^{\circ}$ in decl.). We estimate the uncertainty 
in the angular position of our peak density to 
be $\delta \theta=\sin^{-1}(r'_{k}/r)=1{\circ}.7$ (where $r$ is the distance 
of the density peak and $r'$ is the  radius of the sphere enclosing the fifth 
nearest neighbor of the densest point)--- smaller than the
angular distance of the peak from the SDSS Stripe 82.
Our results favor an interpretation of unbound satellite system or possibly a
bound system within a larger overdensity. 
Such cloud-like structures are expected to be formed from satellites disrupting along eccentric orbits, while the classical rosette tails 
\citep[such as those of the Sagittarius dwarf galaxy, see][]{2003ApJ...599.1082M}
arise from objects on more circular orbits \citep[see][for a more complete discussion of characteristic morphologies]{2008ApJ...689..936J}.

Finally, note that a smaller sub-concentration of RR Lyrae stars, at a median
distance of 92 kpc,
has also been noted in the interval $-25^{\degree}$ to $-20^{\degree}$
of Stripe 82  \citep[structure L in][]{2007AJ....134.2236S}, 
which seems to coincide in angular position (see the upper panel of \fig{f9}) and distance
with a smaller sub-concentration of stars belonging to structure A14
in the M-giant survey (distances estimated to be 95, 80 and 88 kpc for
three M-giants lying in that region). 
Whether A14 and A16 are truly associated
with the structures in RR Lyraes in Stripe 82  (or with each other) can be tested by mapping their velocity and spatial structures.

\section{Summary}\label{sec:summary}

We have explored the use of a density based hierarchical 
clustering algorithm to identify structures in
the stellar halo. Application of the group finder to 
a simulated data demonstrated that in three-dimensional data sets with
large dispersion in the radial dimension, a coordinate
transformation where the radial coordinate is in logarithmic units   
greatly improves the quality of clustering. 

As an application to a real data set we ran the group-finder on
the 2MASS M-giant catalog and
identified 16 structures in it---
10 of these are known structures and six are new. Among the six new
structures, two are probably due to masks employed on the data, one 
is associated with a strong extinction region, and one is probably a
part of the Monoceros ring. Another one 
originates at low latitude, suggesting contamination by disk
stars, but also shows significant protrusions extending to
high latitudes implying that it is a real feature in the stellar halo. 
 
One structure is free from these
defects, has an overdensity similar to that of known structures like the
streams of the Sagittarius dwarf galaxy and is also slightly above the
Poisson noise. While these properties suggest that it is a genuine structure,
possibly a satellite remnant, the structure was 
also found to match a low temperature feature in the dust map. 
The correspondence with a feature in dust map could either mean that
the structure is an artifact of the extinction corrections 
or that the dust feature is associated with the real gas and dust 
in the structure itself. 

The position and distance of the detected structure
closely match those of the Pisces overdensity, which has been recently  
discovered using RR Lyraes in the SDSS Stripe 82.  
If A16 is indeed related to Pisces overdensity then our analysis using 2MASS 
M-giants provides an independent confirmation of the overdensity and
offers an extended view of it. 
In addition, our analysis suggests that the peak point of density is
located just outside the range of the SDSS stripe, which favors the
interpretation that the system is an unbound satellite system,   
probably corresponding to a debris from  a satellite
disrupting along a fairly radial orbit. 
Deeper photometric surveys of this region along with spectroscopic 
measurements of
the giant stars associated with the overdensity should help confirm or 
rule out this scenario.

Overall we conclude that group finding is a promising technique to unravel
the history of our stellar halo and as a window on accretion more generally. 
Clouds of debris like the Pisces overdensity
are naturally found in model stellar halos built within a standard cosmological context, and are even predicted to be the dominant structures in the outer halo \citep{2005ApJ...635..931B,2008ApJ...689..936J}. Indeed, if none were found we would conclude either that we live in a Galaxy that has suffered an unusual paucity of accretion events on radial orbits, or that our expectations of orbital distributions of accreting objects (gleaned from cosmological simulations of structure formation) are flawed. 

Future prospects for group-finding are even brighter:
our analysis here has only used the three-dimensional spatial
distribution of stars while many surveys also have velocity (proper
motions and radial velocities of stars) and chemical abundance information.
These additional dimensions should help recover more structures.  
Moreover, we have here used M-giants as tracers of the stellar halo. 
Since M-giants are metal rich stars 
this means that our sample is  
biased toward high-metallicity systems that originate from high 
mass progenitors and misses out on the much more numerous low mass 
systems that have low metallicity. Hence surveys utilizing a different
tracer population, e.g., main sequence stars or RR Lyraes should
unravel more structures in the stellar halo.

\section*{Acknowledgments}
This project was supported by the {\it SIM Lite} key project {\it Taking
  Measure of the Milky Way} under NASA/JPL contract 1228235. SRM, RRM and JKC
appreciate additional support from NSF grants AST-0307851 and
AST-0807945. 
KVJ thanks Juna Kollmeier, Josh Simon and Ian Thompson for inspiring conversations that contributed towards the final stages of completion of this manuscript   --- and OCIW for hosting her so she could have these conversations.

\providecommand{\newblock}{} 
\bibliographystyle{apj} 
\bibliography{ms}
\end{document}